\documentclass[3p,times]{elsarticle}

\usepackage{ecrc}

\volume{00}

\firstpage{1}

\runauth{V. Papaspirou et al.}

\jid{procs}

\CopyrightLine{2011}{Published by Elsevier Ltd.}

\usepackage{amssymb}

\usepackage[figuresright]{rotating}
\usepackage{color,soul}
\usepackage{color, colortbl}
\usepackage[table]{xcolor}
\usepackage[utf8]{inputenc}
\usepackage[T1]{fontenc}
\usepackage{float}
\usepackage{multirow}
\usepackage{verbatim} 
\usepackage{array}
\usepackage{amssymb}
\usepackage{pifont}
\usepackage{multicol}
\usepackage{caption}
\usepackage{graphicx}
\newcolumntype{P}[1]{>{\centering\arraybackslash}p{#1}}

\begin{document}

\begin{frontmatter}




\title{Cybersecurity Revisited: Honeytokens meet Google Authenticator}







\author[First]{Vasilis Papaspirou\corref{cor1}}
\ead{vpapaspyrou@uniwa.gr}

\author[Second]{Maria Papathanasaki}
\ead{mpapathanasaki@uth.gr}

\author[Third]{Leandros Maglaras}
\ead{leandros.maglaras@dmu.ac.uk}

\author[First]{Ioanna Kantzavelou}
\ead{ikantz@uniwa.gr}

\author[Fourth]{Christos Douligeris}
\ead{cdoulig@unipi.gr}

\author[Fifth]{Mohamed Amine Ferrag}
\ead{ferrag.mohamedamine@univ-guelma.dz}

\author[Sixth]{Helge Janicke}
\ead{helge.janicke@cybersecuritycrc.org.au}

\cortext[cor1]{Corresponding Author}

\address[First]{Department of Informatics and Computer Engineering, University of West Attica, Athens, Greece}
\address[Second]{Department of CS and Telecommunications, University of Thessaly, Lamia, Greece}
\address[Third]{Cyber Technology Institute, De Montfort University, Leicester, UK}
\address[Fourth]{Department of Informatics, University of Piraeus, Piraeus, Greece}
\address[Fifth]{Department of Computer Science, Guelma University, Guelma, Algeria}
\address[Sixth]{Cyber Security CRC and Edith Cowan University, Perth, Australia 6027}


\begin{abstract}

Although sufficient authentication mechanisms were enhanced by the use of two or more factors that resulted in new multi factor authentication schemes, more sophisticated and targeted attacks have shown they are also vulnerable. This research work proposes a novel two factor authentication system that incorporates honeytokens into the two factor authentication process. The current implementation collaborates with Google authenticator. The novelty and simplicity of the presented approach aims at providing additional layers of security and protection into a system and thus making it more secure through a stronger and more efficient authentication mechanism.

\end{abstract}

\begin{keyword}
Honeytoken \sep Authentication \sep Security \sep honeywords \sep 2FA
\end{keyword}

\end{frontmatter}
\begin{multicols}{2}

\section{Introduction} \label{sec_intro}

Communication systems rapidly evolved to meet the cutting edge technologies in the last decade. Remote services, procedures, and tasks have greatly replaced in person ones. Consequently, remote access has proven the use of just one password to protect an online account currently inadequate and insufficient for our needs. Higher level requirements leads us to the development of more sophisticated mechanisms to identify a user, and confront new challenges in nowadays threat landscape \cite{maglaras2019threats}.

Password mechanisms fail to protect a user and the system she uses to an acceptable security level for any number of reasons, two the most important, a) the user is unaware and unable to choose a strong password against brute force and dictionary attacks, if the system will not guide her to do so, and b) password storage, even encrypted in hash format, becomes problematic and improper for the size of the liability it sustains. Passwords act as guards and operating systems establish their protection on passwords. But furthermore, users share without hesitation their passwords to facilitate cooperation, to help friends, colleagues or teammates, and do not realize the consequences of such a practice. In other cases, a user might fall victim to a phishing attack and unintentionally discloses her credentials.

In most of the aforementioned cases, a really high risk threat for any computing device takes place, the impersonation of a user, the well known \textit{masquerade attack}. The rationale behind this is that a single password does not guarantee any more the security of an online account and does not protect the system of unauthorized access. Nevertheless, user authentication remains at the first line of defense against such threats, and it is the central component of any security infrastructure. Therefore, enhanced security requires several levels of protection towards an in depth security and defense. In case the first line of defense fails, there is always a way to avoid a possible attack, through a multi-factor authentication (MFA) mechanism.

Multi Factor Authentication (MFA) is achieved by using two or more different factors related to the user \cite{NIST2021}. Such factors may refer to something the user knows, as a password or a PIN. Another factor is connected with  something the user has in his possession, as an identity or a smart card. Finally, another group of factors might be something the user is, any physical feature as a fingerprint, the iris, or the voice, etc. can be used as a method of identification \cite{rathgeb2010two}. A prerequisite for a secure two factor authentication mechanism (2FA) is the use of two separate devices for entering at least two factors \cite{NIST2021}. Unfortunately, this is not always the practice used, which generates one of source of problems in 2FAs. Other problems on the functioning of a 2FA scheme, instead of securing more a users' account, create additional security concerns and makes it even more vulnerable \cite{Certic2018}. SIM Swapping, one of the major recent threats, has been proven to be very difficult to be stopped. Mobile Network Operators (MNOs), banks and authorities have started collaborating to mitigate fraudulent SIM swapping through the use of Application Programming Interface (API) provided by the MNOs to check whether a SIM swap has been recently performed. Even though this solution could be effective against SIM Swapping it is not yet used by all banks and it is not efficient against another family of attacks. The apps a user downloads may not be the only things which are on his/her phone. Hidden spyware that allows people to monitor activity, access information and even eavesdrop on user's chats can also be installed. These malware are called 'Stalkerware' and pose a major threat to mobile devices.

In this paper, we use  the state-of-the-art solutions and propose a novel system that integrates honeytokens in a two factor authentication scheme that is embedded into google authenticator. The novel approach aims at providing additional layers of defense into the system and thus making it more secure through a stronger and enhanced authentication mechanism. 

The contributions of the proposed research work include:
\begin{itemize}
    \item The incorporation of the honeywords technique used as an additional factor on a different computing device to enhance  security,
   \item The use of the very popular QR-code system to accelerate the identification procedure.
    \item The integration and harmonization with the main-stream authentication technology that is widespread today.
    \item The proposal of new integrated mechanism entitled 'Two Factor Honeytoken Authentication Mechanism'
    \item The use of any authenticator, with the Google authenticator used for the pilot system presented in this article.
    \item The low complexity, the simplicity and the user friendliness of the proposed system, even to the elderly.
    \item The security evaluation of the proposed system that is proven to be robust against many serious attacks like SIM swapping, stalkerware and side channel attack among others.
\end{itemize}

The paper is organized as follows.  Section \ref{sec_auth} gives details on Authenticators along with a comparative analysis. 
In Section \ref{sec_prop} we thoroughly describe the proposed system. In Section \ref{sec_sec} we perform a security analysis of the proposed 2FHA mechanism. In Section \ref{sec_related}, we present other research works related to user authentication and 2FA. Finally, in Section \ref{sec_concl} we summarize the findings of our work by writing up conclusions and present possible work extensions that will expand the research outcomes.

\section{Authenticators Comparison}\label{sec_auth}

There is a growing range of Authenticators. Among the Authenticators that are flooding the Internet, there are some that have stood out for the simplicity and efficiency they offer \cite{aggrawal2012authentication}. Google Authenticator (GA) is the most popular as it offers a large number of advantages. It is free and very easy to use. It does not require an Internet connection and supports Time-based One-Time Password (TOTP) and HMAC-based One-Time Password (HOTP). There is a small number of disadvantages including the inability to create a backup and the lack of a variety of features \cite{de2013comparative}. 

Authy 2Factor Authenticator (Authy 2FA) is a widely used Authentication method that has a lot in common with Google Authenticator. It is free and easy to use, with many features and supports crypto-wallets and backup creation. Despite the many benefits it offers, the Multi-device synchronization feature raises security risks since the use of a password in the App is not mandatory, making it difficult to maintain control on all devices. Also, Authy continues to use SMS as an Authentication method while it is generally considered obsolete in 2FA systems. It is worth noting that users of other Authentication applications who wish to transfer all their tokens to Authy will need to have a rooted phone, exposing the overall security of the device to major risks \cite{polleit2018defeating}.

As for Microsoft Authenticator (MA), it is especially useful for users of Microsoft Services, those who have a Microsoft account or have the Windows 10 Operating System on their mobile phone. A great advantage of using this application, is that it notifies the user in case the application is used in an unknown environment and supports passwordless authentication with Microsoft apps. The relatively difficult user interface of the application reduces its popularity, and the fact that it lacks features to make it appealing \cite{polleit2018defeating}.

All the above characteristics are summarized in the following table (Table 1), where \ding{51} represents the existence of a feature in the authenticator and the \ding{55} represents its absence. 
\begin{table}[H]
    \begin{tabular}{ p{4.5cm}  p{0.6cm} p{0.6cm} p{0.4cm} }
     \hline
     \textbf{FEATURES}& \textbf{GA}     & \textbf{Authy 2FA} & \textbf{MA}\\
     \hline
     Open source                        &   \ding{55}   &   \ding{55}   &   \ding{55}\\
     Free                               &   \ding{51}   &   \ding{51}   &   \ding{51}\\
     Widely adopted                     &   \ding{51}   &   \ding{51}   &   \ding{55}\\
     Lack of features                   &   \ding{55}   &   \ding{55}   &   \ding{51}\\
     Easy to use                        &   \ding{51}   &   \ding{51}   &   \ding{55}\\
     Data backup                        &   \ding{51}   &   \ding{51}   &   \ding{51}\\
     Network connection needed  &   \ding{55}   &   \ding{55}   &   \ding{55}\\
     Multiple account support           &   \ding{51}   &   \ding{51}   &   \ding{51}\\
     Cryptocurrency securing    &   \ding{51}   &   \ding{51}   &   \ding{51}\\
     Microsoft services compatible      &   \ding{55}   &   \ding{55}   &   \ding{51}\\
     TOTP and HOTP use                  &   \ding{51}   &   \ding{51}   &   \ding{51}\\ \hline
    \end{tabular}
    
\caption{Most used authenticators comparison}
    \label{tab:my_label}
\end{table}

Authentication mechanisms use cryptographic algorithms to protect user credentials; however, many users have a tendency to choose weak passwords, i.e., common words that can easily be guessed by a dictionary attack \cite{genc2017examination}. The hash value of a frequent password (i.e., a word in a dictionary) may be cracked rapidly thanks to advances in GPU technology. An adversary may be able to get user credentials as a result of these assaults.




\subsection{Honeywords}

Juels and Rivest \cite{juels2013honeywords} advocated the use of "honeywords". The general idea of honeywords is to alter the location where the passwords are stored in a way that each user has a password and some phonies passwords. Honeywords are the passwords that are fake and the sum of all the Honeywords and the password are named Sweetwords. When a honeyword is send by the login phase, there will be an alert that the password database has been infringed.
A set of false passwords is mixed with the user's genuine password, and the hash values of these passwords (real password and honeywords) are kept in the password file to identify whether the password file has been stolen or not.
Even if this file has been hacked and all of its hash values have been broken, the adversary still has no idea which one is the real password.

\section{Proposed System}\label{sec_prop}

In this paper, we suggest the use of an alternative method, the 'Two Factor HoenyToken Authentication (2FHA)' mechanism. We propose the combination   Honeywords and two factor authentication in a novel  mechanism. The mechanism collaborates with Google authenticator.

This mechanism agrees with a new user two factors, a password and the right position of a honeyword it will provide to him at log in on a different device (eg. a mobile phone). This covers the 2FA concept and upon login the user has to enter two pieces of information, the password and the correct honeyword. The user knows where the correct honeyword has been placed in the sequence of honeywords an authenticator sends to him on the second device. Then, he enters it to complete his login to the system.

A new user first has to get registered through a simple registration form 
During this procedure, the user is prompted to enter a number from a specific range (i.e. 1 to 3). This number is agreed between the mechanism and the user and indicates the correct position where the genuine OTP number will be placed, among a sequence of fake OTP passwords, during the login phase. A new sequence of fake OTP passwords will be produced each time the user attempts to login, and will include the genuine OTP number at the right position. The length of this sequence has been chosen to be $3$ for the pilot implementation. Detailed operational descriptions on the registration and login phases are provided in the following sections.


\begin{figure}[H]
    \centering
    \includegraphics[width=0.45\textwidth]{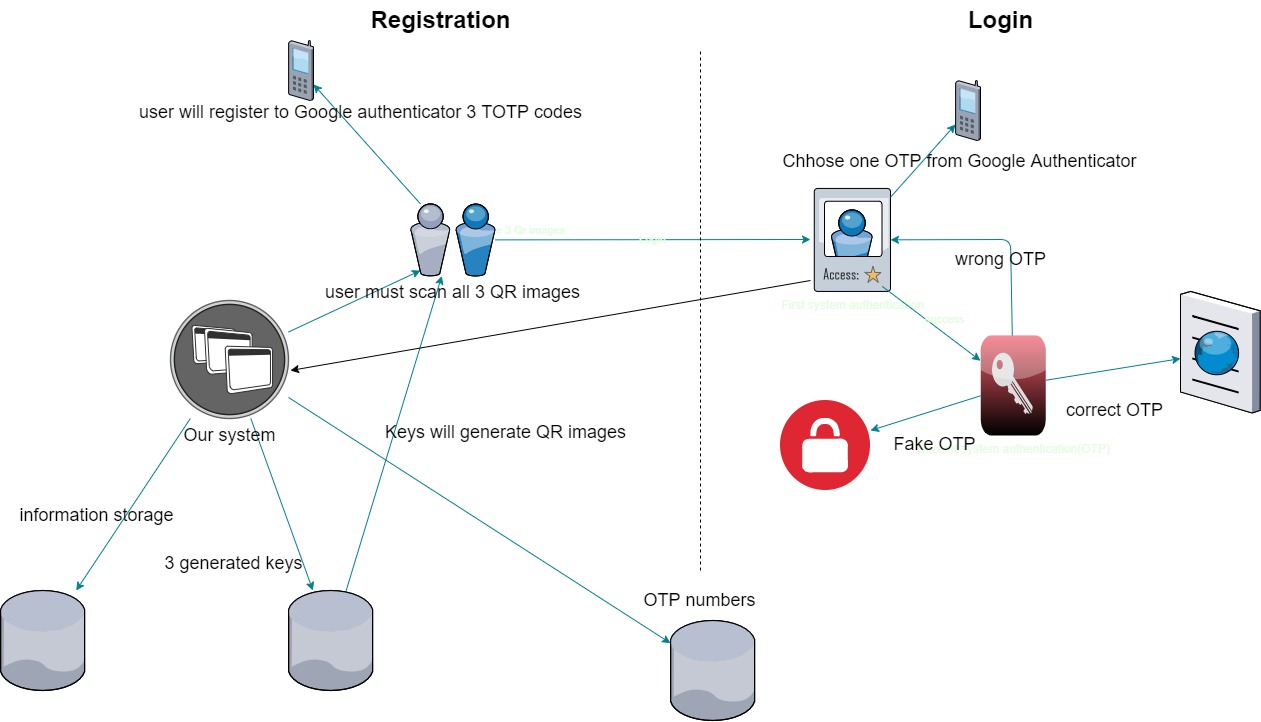}
    \caption{The architecture and functioning of the proposed system}
    \label{fig:our system}
\end{figure}

{\bf Registration phase}\\

A user must enter first some information into the proposed mechanism, in order to get registered. He has to fill the required fields (username, password, firstname, lastname, phone) to be identified at the login phase later 
The phone field requires the phone number,  which is the device where the OTP codes will be sent at the login phase. Afterwards, the user has to choose the position of the right OTP code, a number from a predetermined range of numbers available to him $(1,2,3)$. The number the user selects will be the place the correct OTP code will appear in the sequence of the three OTP codes. For instance, if he chooses number 2, then from the three OTP codes he will receive in the login phase, number 2 will be the correct one. The username must be unique.


When the user is registered, the system will produce three keys, which are not the same in length. These keys will generate the OTP codes for the user. If the registration is successful, the user will receive three QR-Images that must be scanned with the Google Authenticator. The QR-images are compatible with other authenticators as well, like Authy.
The database of the system will not save any OTP codes.The OTP number the user chooses, will not be stored in the same database with the rest of the information. For security reasons they will be saved in another database.

{\bf Login phase}

At the login phase, a user will have to complete the username and the password fields, and this is the first layer of security in our system. 
If the user enters three times a wrong password in a row, his account will be locked. When the user successfully enters these credentials into the system, he will receive an SMS with three OTP codes. In addition, the user is authorized to access the OTP codes via the Google authenticator. When the user receives the three OTP codes from the authenticator or via SMS, he knows which one to select because he has chosen the correct one during the registration phase. By picking the correct number, he will be able to fill the OTP edit box in successfully. If the user selects any other OTP code, then the account will be locked and the user will be informed that an attempt of breach occurred. 


\section{Security Analysis}\label{sec_sec}

In this model we focus our research at the security of the user authentication mechanism using honeywords. Without increasing login complexity  and without loss of the general idea of the honeywords, the same key factors remain in our system. In this section we  discuss several attacks and how our system is robust to them.


{\bf Stalkerware}

The apps a user downloads may not be the only things which are on his/her phone. Hidden spyware that allows people to monitor activity, access information and even eavesdrop on user's chats can also be installed. 2FHA mechanism is robust to stalkerware since even the attacker manages to read the OTPs that are send to the mobile phone of the user he will not be able to know which is one is the valid one among the fake ones.

{\bf SIM Swapping}

Since 2017 there have been several media reports about SIM swapping attacks, targeting people within the cryptocurrency community, but also bank accounts and social media and email accounts. SIM Swapping can be used to bypass 2FA security but our proposed 2FHA mechanism is robust against this attack, since only the legitimate user knows the position of the valid OTP among the fake ones.

{\bf Video recording Attack}

This attack is using a video recording device as an external tool for the attack. At most cases it is very difficult to make this attack without the user's noticing it. On the one hand, the use of username and password, and on the other hand the inability of the attacker to scan the QR-images that would grant him access to Google Authenticator, make our proposed system robust to such attack. 

{\bf Guessing Attack}

The success of the guessing attack depends on the length of the password.The larger the length, the lower the chance of a successful attack.  In our system the attacker will have to guess not only the first layer of defence (username, password) but also the second one (OTP).  Even if he gets access through the first layer, he must be aware of the correct OTP number to choose. In the worst scenario he must choose one of the three OTP codes that are shown to him. If he chooses the wrong one, the account will be locked.

{\bf Side Channel Attack}

These attacks sometimes require the installment of a software (trojan horse) to the user's computer, to extract the information they want.
In the case of the proposed mechanism, even if the attacker gains access to the username and password, he will be asked to get through the second security layer, which requires the input of the correct OTP , in order to gain access to the system.

We analyzed several attacks that our system could face. As we have seen, the proposed mechanism is robust to all these attacks since in order for an attacker to gain access to a user account, he must successfully pass all the security layers we have created. On the one hand, this process is very complicated for the attacker and on the other hand it doesn't complicate the authentication process for the user.

\section{Related Work}\label{sec_related}

Recently, three techniques for completely removing the possibility of offline password guessing have been proposed: (a) using a machine-dependent function (e.g., Ersatz Passwords \cite{almeshekah2015ersatzpasswords}); (b) using distributed cryptography (e.g., threshold password-authenticated secret sharing \cite{camenisch2015optimal}); and (c) using external password-hardening services (e.g., Phoenix \cite{lai2017phoenix}). All these techniques, however, need significant modifications to server-side authentication mechanisms. In addition, because they have a lack of scalability, technique (a) does not enable backing up password files where hash values are kept in a distributed way, making the mechanism inappropriate for Internet-scale services. Technique (b) necessitates client-side system changes, which is not user-friendly and widely admitted as not desirable; Technique (c) is subject to a single point of failure and may leak user behavior information to external parties. Many users have a tendency to choose weak passwords, i.e., common words that can be easily guessed by a dictionary attack [\cite{florencio2007large},\cite{furnell2000authentication}], when their choices are not monitoring and guided by the authentication mechanism itself.

In the literature, the use of honeywords for password protection was first proposed by A. Juels and R. Rivest, in 2013. Later, many works were based on this paper, others expanded it and others improved its vulnerabilities. Honeywords have also been suggested for use in authentication, however,  in this paper we recommend their use in a novel two factor authentication scheme.

A. Juels and R. Rivest \cite{juels2013honeywords} proposed a method to make the security of hashed passwords more powerful. They used honeywords to make password cracking detectable.
More specifically, an attacker who achieves to access a file that  contains fragmented passwords, inverting the hash function, is unable to distinguish the real ones from honeywords. If the attacker tries to enter a honeyword, a "honeychecker", which knows what the honeywords are, is activated and warns for an intrusion attempt.

In \cite{wang2016two} a new, highly efficient system is proposed, which aims to resolve issues arising from user corruption and server compromisation. Honeywords are adopted to eliminate the long-standing security-usability conflict, and they do user and sensor authentication to resist the node from becoming compromised. 

In \cite{10.1007/978-3-319-93524-9_8} the authors extend the work of A. Juels and R. Rivest by covering some of its gaps. They suggest a solution to the possibility an attacker to modify the honeychecker code, or the code executed on the login server, thus activating the alarm.

In \cite{10.1007/978-3-030-24643-3_105} the work of A. Juels and R. Rivest is coming to the forefront again. The authors here, however, use honeywords not only to protect passwords, but also to detect potential intruders on the system. It is important that they do not just detect, but continue to gather information about the malicious intruder, tracking the source of the problem in the system.

Finally, in \cite{li2019two} the authors conclude that some standard, present authentication schemes intended to authenticate that are based on IIoT, do not meet all the security requirements set by modern era. For this reason, they created a new authentication scheme, much more powerful than the existing ones, which adopts the honeywords technique. This scheme suggests a way for effective detection and thwart node capture attack.

The proposed work presented in this article attempts to enhance a traditional 2FA mechanism in order to overcome security problems related to password authentication mechanisms, 2FA mechanisms, and users' problematic practices and behaviors that finally endangers heavily computing systems. The selected approach aims at protecting a system and its users from unauthorized access to resources, from data leakage and disclosure of private information, and provides secure and uninterruptible services to the next possible level. The Honeywords principles are used to detect corrupted or stolen tokens. By introducing also QR codes, the integration of this mechanism into any platform or web application and gain access via a mobile phone can be easily accomplished. This work has been developed over the previously published one, which outlines the main concept \cite{papaspirou2021novel}.

\section{Conclusions and future work}\label{sec_concl}

We have analyzed the mix of Honeywords with a two factor authentication mechanism in our system that is entitled 2FHA. Our proposed mechanism increases the complexity for the attacker, is robust to SIM Swapping attacs an long as to a number of other well known attacks.

The mix system we propose, has the advantage that it goes hand by hand with the mainstream authentication technology, most users have nowadays. 
Moreover,  Google authenticator is compatible  with the systems of many companies that use already for security reasons.  On the other hand, all these security requirements impose additional cost to the users. In order for a user to access Google authenticator, he must own a smartphone which is not affordable by everyone. Also, due to the fact that the elderly are less familiar with the use of smart  applications, the proposed 2FHA mechanism supports the option of sending simple SMSs in order to facilitate them.

For the future we will try to upgrade the defense system and to make it platfrom agnostic.  We plan to test the proof of concept by integrating the 2FHA mechanism to banking or health organizations. These are the main targets  of recent cyber-attacks and since they are already using OTP systems for online transactions the proposed 2FHA  could enhance their security.  

\section{Acknowledgement}

This work has received funding from the European Union’s Horizon 2020 research and innovation programme: projects CyberSec4Europe (Grant Agreement no. 830929), and LOCARD (Grant Agreement no. 832735)

\bibliographystyle{elsarticle-num}
\bibliography{ictexpress.bib}
\end{multicols}
\end{document}